\documentclass{aa}
\usepackage{graphics}
\usepackage{epsfig}

\def\mearth{M_\oplus}
\def\msun{M_\odot}

\def\rhill{R_{\rm Hill}}
\def\racc{R_{\rm acc}}

\def\aplanet{a_{\rm planet}}
\def\mplanet{M_{\rm planet}}

\def\racc{R_{\rm accr}}
\def\mcore{M_{\rm core}}

\def\mearth{M_\oplus}
\def\aini{a_{\rm start}}

\def\mdotst{\dot{M_{st}}}
\def\flambda{f_{\Lambda}}
\def\sigmaP{\Sigma_{\rm P}}
\def\mcore{M_{\rm core}}

\def\aplanet{a_{\rm planet}}
\def\Mplanet{M_{\rm planet}}
\def\rhill{R_{\rm H}}

\def\msol{M_\odot}

\def\simgr{\,\hbox{\hbox{$ > $}\kern -0.8em \lower 1.0ex\hbox{$\sim$}}\,}
\def\simle{\,\hbox{\hbox{$ < $}\kern -0.8em \lower 1.0ex\hbox{$\sim$}}\,}
\def\beq{\begin{equation}}
\def\eeq{\end{equation}}
\def\dpartial#1#2{{{\partial {#1}} \over {\partial {#2}}}}

\def\msol{M_\odot}
\def\Mstar{M_*}

\def\simgr{\,\hbox{\hbox{$ > $}\kern -0.8em \lower 1.0ex\hbox{$\sim$}}\,}
\def\simle{\,\hbox{\hbox{$ < $}\kern -0.8em \lower 1.0ex\hbox{$\sim$}}\,}
\def\beq{\begin{equation}}
\def\eeq{\end{equation}}
\def\dtotale#1#2{{{d {#1}} \over {d {#2}}}}

\def\aj{AJ}                   
\def\araa{ARA\&A}             
\def\apj{ApJ}                 
\def\apjl{ApJ}                
\def\apjs{ApJS}               
\def\ao{Appl.Optics}          
\def\aap{A\&A}                
\def\mnras{MNRAS}             

\def\({\left(}
\def\){\right)}
\def\<{\left<}
\def\>{\right>}

\begin{document}

\title{Models of Giant Planet formation with migration and disc evolution}

\author{Yann Alibert \inst{1} \and Christoph Mordasini \inst{1} \and Willy Benz \inst{1} \and Christophe Winisdoerffer \inst{2}}

\institute{
$^1$ Physikalisches Institut, University of Bern, Sidlerstrasse 5, CH-3012 Bern, Switzerland \\ 
$^2$ Theoretical Astrophysics Group, University of Leicester, Leicester,
LE1 7RH, United Kingdom
} 

\offprints{ Yann ALIBERT, \email{yann.alibert@phim.unibe.ch}}

\date{Received 20 september 2004 / Accepted 13 december 2004}

\abstract{
We present a new model of giant planet formation that 
extends the core-accretion model of Pollack {\it et al.} (\cite{P96})
to include migration, disc evolution and gap formation.
We show that taking into account these effects can lead
to a much more rapid formation of giant planets, making
it compatible with the typical disc lifetimes inferred
from observations of young circumstellar discs.
This speed up is due to the fact that migration prevents
the severe depletion of the feeding zone as observed in 
{\it in situ} calculations. Hence, the growing planet is 
never isolated and it can reach cross-over mass on a much
shorter timescale. To illustrate the range of planets that
can form in our model, we describe a set of simulations in
which we have varied some of the initial parameters and 
compare the final masses and semi-major axes with those
inferred from observed extra-solar planets.

}

\titlerunning{Giant planet formation}
\authorrunning{Alibert Y. et al.}
\maketitle

\section{Introduction}

The standard giant planet formation scenario is the so-called core-accretion 
model. In this model, a solid core is formed first by accretion 
of solid planetesimals which themselves were formed by sedimentation and 
coagulation of small dust grains (Wetherill \& Steward 1989, Lissauer 1993). 
As the core grows, it eventually becomes massive enough to gravitationally 
bind some of the nebular gas thus, surrounding itself by a tenuous envelope. 
The subsequent evolution of this core/envelope structure has been studied 
first by Perri \& Cameron (1974) and subsequently in great detail by 
Bodenheimer \& Pollack (1986 hereafter referred to as BP86) and Pollack 
{\it et al.} (1996 hereafter referred to as P96). These authors have found 
that both the solid core and the gaseous envelope subsequently grow in mass, 
the envelope remaining in quasi-static and thermal equilibrium. During this 
phase, the energy radiated by the gas is supplied by the energy released 
from the accretion of planetesimals. As the planet reaches a high enough 
mass (of the order of $20-30 \mearth$ at $5$ AU,  but depending on different 
physical parameters such as the solid accretion rate), radiative 
losses become so large that they can no longer be offset by planetesimal 
accretion alone and the envelope starts to contract. At this stage,
the mass is nearly equally distributed between the mass of accreted gas,
and the mass of accreted
planetesimals.  The contraction of the envelope 
increases the gas accretion rate which in turn increases the energy losses 
through radiation and the process runs away very rapidly building up a 
massive envelope.

Detailed numerical calculations in particular by P96 have shown that this 
core-accretion formation model can be subdivided in three distinct phases. 
During phase 1, the core is formed by accretion of planetesimals located 
inside the growing planet's feeding zone which extends to a few Hills radii. 
The accretion rate of solids during this phase is typically of the order 
of $10^{-5} \mearth /$yr while the gas accretion rate is several orders 
of magnitude lower. Phase 1 ends when the feeding zone of the planet becomes 
severely depleted, which generally occurs before the planet has reached a
mass high enough to accrete a large amount of gas. 
During phase 2, the mass increase is essentially due to the 
slow accretion of gas in the envelope. Note that by increasing the envelope
mass, the feeding zone of the planet also increases which in turns allows 
the accretion of more planetesimals. Both accretion rates (solid and gas) 
turn out to be relatively constant during this phase with the gas accretion 
rate exceeding by a fraction of an order of magnitude the solid accretion 
rate. This phase lasts until the planet has reached the cross-over mass
(mass of accreted planetesimals  equal mass of accreted gas),
at which time the system enters phase 3. At this point, evolution proceeds 
extremely fast by runaway gas accretion as the envelope is no longer
able to maintain quasi-static equilibrium. The mass of the planet increases 
correspondingly.  The timescale for the formation of a giant planet in this 
core-accretion scenario is almost entirely determined by phase 2, which, 
as shown by P96, turns out to be an extremely sensitive function of the 
disc surface density. P96 found for their preferred model for the formation 
of Jupiter (model J1, the surface density being equal to $\sim 4$ times the 
value in the minimum mass solar nebula) a formation timescale close to 8 Myr 
while reducing the surface density to 75\% of this value leads to a 
formation time of nearly 50 Myr.

The core-accretion scenario has been motivated by the apparent existence of 
a solid core, long estimated to be of the order of the mass of accreted
planetesimals when the planet reaches the cross-over mass, in 
all the giant planets of the solar system. This scenario has also been supported
by the  enrichment in heavy 
elements (heavier than He) of both Jupiter and Saturn compared to solar 
value, deduced from interior structure models and various remote/in situ 
measurements (radius, mass, surface abundance, gravitational moments, see 
Guillot et al. 2004).

However, the major difficulty affecting the core-accretion scenario which 
has been pointed out repeatedly is related to the timescale required to form 
a giant planet. Based on astronomical observations, protoplanetary discs are 
believed to transport mass inward at a rate of $10^{-8 \pm 1} \msol /$yr 
(Hartmann {\it et al.} 1998) while their total mass has been estimated to 
lie between $10^{-3}$ and $10^{-1} \msol$ (Beckwith \& Sargent 1996). From 
these numbers as well as from the observations that about half the stars in 
young clusters loose their discs within 3 Myr (Haisch {\it et al.} 2001) one 
infers a typical circumstellar disc lifetime of  1-10 Myr which is of the 
same order to (if not smaller than) the giant planet formation timescale. To 
circumvent this problem, Boss (2002) has proposed that giant planets form 
directly from the gravitational fragmentation and collapse of a protoplanetary
disc. However, there are still a number of open issues about this scenario 
such as the formation and survival of bound structures since most calculations 
so far have used an isothermal equation of state and/or too low resolution. 
Furthermore, while solving the timescale problem, it is not clear at all if 
the peculiar composition and structure of Jupiter and Saturn can be explained 
within this model. More work is definitively required to investigate this 
scenario further.

Since the discovery by Mayor \& Queloz (1995) of the first extrasolar giant 
planet at short distance to its star, there is mounting evidence that planets 
might have formed at locations which do not necessarily correspond to those 
where they are observed today. Gravitational interactions between the growing 
planet and the gaseous disc lead to angular momentum transfer resulting in a 
net inward migration of the planet and possibly gap formation (Lin \& 
Papaloizou 1986, Lin {\it et al.} 1996, Ward 1997, Tanaka {\it et al.} 2002).
 For low mass planets, the migration rate is linear with mass (type I migration, 
Ward 1997). Higher mass planets open a gap and the migration rate is set by 
the viscosity independently of planetary mass (type II migration, Ward 1997). 
While the general physical understanding of the origin of migration is clear, 
the actual migration rates obtained, especially for type I migration, are 
so short that they are inconsistent with the number of extrasolar planets 
actually detected. Type II migration timescales are found to lie between 
0.1 to 10 Myr, a timescale comparable to (if not shorter than) the typical
lifetime of the disc as well as the planet formation timescale in the 
core-accretion scenario.

Since all relevant timescales (planet formation, disc evolution, and migration) 
are of the same order of magnitude, it appears difficult to obtain a 
self-consistent picture while omitting anyone of these processes. For this 
purpose, we have developed a new code, structured around three modules. These 
modules calculate the disc structure and evolution (using the simple formalism 
of $\alpha$ viscosity), including planet migration, the interaction of 
planetesimals with the planet atmosphere, and the planet structure and 
evolution. In section 2 we will describe these modules in more details and 
present the numerical tests we have performed to validate them. Section 3 
will be devoted to the effect of migration and disc evolution on formation 
timescales, and to formation models of extrasolar planets. Finally, we 
will discuss these results and conclude in section 4.

\section{Equations and assumptions}

Our code to compute the formation of giant planets consists of three modules. 
The first module follows the method given by Papaloizou \& Terquem (1999, 
referred to PT99 in the following) to calculate the disc structure and its 
time evolution. The second one calculates the interaction between the 
infalling planetesimals and the atmosphere of the growing planet; the last 
one is a  planetary structure and evolution code, written especially for 
this project. We now describe these modules, and present some of the tests 
we have performed.

\subsection{Disc structure and evolution}
\label{disk}

\subsubsection{Vertical structure}
\label{disk_vert}

The first module aims at determining the structure (both vertical and 
radial) of a protoplanetary disc. The method is identical to the one used 
by PT99 and therefore we only briefly recall the main points. Cylindrical 
symmetry is assumed and therefore the cylindrical coordinates $(r,z,\theta)$ 
are a natural choice. The disc is assumed to be thin. For each distance $r$
to the star, the vertical structure is calculated by solving the equation 
of hydrostatic equilibrium,
\beq
{1 \over \rho}  \dpartial{P}{z} = - \Omega^2 z
\label{eq_disk_hydro}
,
\eeq
where $z$ is the vertical coordinate, $\rho$ is the density and $P$ is 
the pressure. The disc is assumed to be Keplerian with $\Omega^2 = G 
\Mstar / r^3$, $G$ being the gravitational constant and $\Mstar$ the mass 
of the central star, assumed to be equal to the solar one.  This equation is 
solved together with the energy equation:
\beq
\dpartial{F}{z} = {9 \over 4} \rho \nu \Omega^2
\label{eq_disk_ener}
,
\eeq 
which states that the energy production by the viscosity $\nu$ is removed 
by the radiative flux $F$, and the diffusion equation for the radiative flux:
\beq
F = {- 16 \pi \sigma T^3 \over 3 \kappa \rho} \dpartial{T}{z}
\label{eq_disk_diff}
,
\eeq
where $T$ is the temperature, $\kappa$ is the opacity, and $\sigma$ is 
the Stefan-Boltzmann constant.  The viscosity is calculated using the standard 
Shakura \& Sunyaev (1973) $\alpha-$parametrization $\nu = \alpha C_s^2 / 
\Omega$ where the speed of sound $C_s^2$ is determined from the equation 
of state.

The boundary conditions for this part of the calculation are the same as 
in PT99, formally,
\beq
T(z=H) = T(\tau_{\rm ab},T_b,r,\mdotst,\alpha)
,
\eeq
\beq
P(z=H) = {\Omega^2 H \tau_{\rm ab} \over \kappa(T(z=H),P(z=H))}
,
\eeq
\beq
F(z=H) = {3 \over 8 \pi} \mdotst \Omega^2
,
\eeq
and
\beq
F(z=0) = 0
.
\eeq

These conditions depend on three parameters: $\tau_{\rm ab}$ the 
optical depth between the surface of the disc ($z = H$) and infinity, 
$T_b$ the background temperature, and $\mdotst$ the equilibrium 
accretion rate defined by $\mdotst \equiv 3 \pi \tilde{\nu} \Sigma$ 
where $\Sigma \equiv \int_{-H}^{H} \rho dz$ is the usual surface 
density, and $\tilde{\nu} \equiv \int_{-H}^{H} \nu \rho dz / \Sigma$. 
The values for $\tau_{\rm ab}$ and  $T_b$ are the same as in PT99 
(namely $10^{-2}$ and $10$ K); the steady-state accretion rate
is a free parameter. As shown in PT99, the structure obtained hardly 
varies with the first two parameters.

This system of 3 equations with 4 boundary conditions has in general 
no solution, except for a certain value of $H$. This value is found 
iteratively: equations \ref{eq_disk_hydro} to \ref{eq_disk_diff}
are numerically integrated from $z=H$ to $z=0$, using a fifth-order 
Runge-Kutta method with adaptive step length (Press {\it et al.} 1992) 
until $F(z=0)=0$ to a given accuracy. 

Using this procedure, we calculate, for each distance to the star $r$ 
and each value of the equilibrium accretion rate $\mdotst$, the 
distributions of pressure, temperature and density $T(z;r,\mdotst)$, 
$P(z;r,\mdotst)$, $\rho(z;r,\mdotst)$.

Using these distributions, we finally calculate the mid-plane temperature 
($T_{mid}$) and pressure ($P_{mid}$), as well as the effective viscosity 
$\tilde{\nu}(r,\dot{M_{st}})$, the disc density scale height $\tilde{H}
(r,\dot{M_{st}})$ defined by $\rho(z=\tilde{H}) = e^{-1/2} \rho(z=0) $.
The surface density $\Sigma(r,\dot{M_{st}})$ is also given as a function of
$\mdotst$ (for each radius). By inverting this former relation,
we finally obtain relations $T_{mid}(r,\Sigma)$, $P_{mid}(r,\Sigma)$, 
$\tilde{\nu}(r,\Sigma)$ and $\tilde{H}(r,\Sigma)$ for each value of $r$ 
(and each value of the other parameters $\alpha$, $\tau_{\rm ab}$ and $T_b$).

 
\subsubsection{Evolution of the surface density}

The time evolution of the disc is governed by the well-known diffusion 
equation (Lynden-Bell \& Pringle 1974):
\beq
{d \Sigma \over d t} = {3 \over r} {\partial \over \partial r } \left[ r^{1/2} {\partial \over \partial r}
( \tilde{\nu} \Sigma r^{1/2})  \right] = {1 \over r} {\partial \over \partial r } \left( r J(r) \right)
\label{eqdiff_std}
,
\eeq
where $J(r) \equiv {3 \over r^{1/2} } {\partial \over \partial r }(\tilde{\nu} 
\Sigma r^{1/2})$ is the mass flux (integrated over the vertical axis $z$).
This equation is modified to take into account the momentum transfer between 
the planet and the disc, as well as the effect of photo-evaporation and accretion 
onto the planet: 
\beq
{d \Sigma \over d t} = {3 \over r} {\partial \over \partial r } \left[ r^{1/2} {\partial \over \partial r}
\tilde{\nu} \Sigma r^{1/2} + \Lambda (r) \right] + \dot{\Sigma}_w(r) + \dot{Q}_{\rm planet}(r)
\label{eqdiff}
.
\eeq
The rate of momentum transfer $\Lambda$ between the planet and the disc is 
calculated using the  formula derived by Lin \& Papaloizou (1986):
\beq
\Lambda(r) ={ \flambda \over 2 r}\sqrt{GM_{\rm star}}\left( {M_{\rm planet} \over M_{\rm star}} \right)^2
 \left( {r \over {\rm max}( |r-a| , \tilde{H} )} \right)^4
,
\eeq
where $a$ is the sun-planet distance and $\flambda$ is a numerical 
constant\footnote{In this formula, 
the disc scale height $\tilde H$ is the scale height of the unperturbed disc,
and not the scale height in the middle of the gap.}.
The photo-evaporation term $\dot{\Sigma}_w$ is given by (Veras \& Armitage 2004):
\begin{eqnarray}
\left\{
\begin{array}{l}
\dot{\Sigma}_w  = 0  \,\,\,\,\,\,\,\,\,\,\, {\rm for}  \,\,\, R < R_g , \\
\dot{\Sigma}_w  \propto R^{-1}  \,\,\, {\rm for}  \,\,\, R > R_g ,
\end{array}
\right.
\end{eqnarray}
where $R_g$ is usually taken to $5$ AU, and the total mass loss due to 
photo-evaporation is a free parameter. Finally, a sink term $\dot{Q}_{\rm planet}$ 
is included in Eq. \ref{eqdiff}, to take into account the amount of gas accreted 
by the planet. This term is generally negligible compared to the other ones, 
except during the runaway phases.

To solve the diffusion Eq. \ref{eqdiff} we need to specify two boundary 
conditions. The first one is given at the outer radius of the disc (in our 
simulations this radius is usually taken at $50$ AU). At this radius, one can 
either give the surface density $\Sigma$ or its temporal derivative. Since 
the characteristic evolution time of the disc is the diffusion timescale
\beq
T_{\nu} \propto {r^2 \over \tilde{\nu}} \propto {1 \over \alpha \Omega } \( {r \over H} \)^2 
\label{eq_Tnu_disk}
,
\eeq
which\footnote{The second part of Eq. \ref{eq_Tnu_disk} is obtained by 
expressing the equation \ref{eq_disk_hydro} as ${1 \over \rho} {P \over H} 
\sim \Omega^2 H$ and then replacing the sound velocity by $\Omega H$ in the 
definition of $\nu$.} is proportional to $r^{3/2}$ for discs of approximately 
constant aspect ratio (which is the case in these models (see PT99)) the 
outer boundary condition has little influence.

The second condition is specified at the inner radius where we have used 
the following condition:
\beq
 r \dpartial{ \tilde{\nu} \Sigma}{r} \bigg|_{\rm inner \,\,\, radius} = 0
\label{BC_disk_int}
.
\eeq 
Since the total mass flux through a cylinder of radius $r$ is given by:
\beq
\Phi(r) \equiv  2 \pi r J(r) = 3 \pi \tilde{\nu} \Sigma + 6 \pi r \dpartial{\tilde{\nu} \Sigma}{r}
,
\eeq
the boundary condition Eq. \ref{BC_disk_int} can be expressed as:
\beq
\Phi(r)  \Big|_{\rm inner \,\,\, radius} = 3 \pi \tilde{\nu} \Sigma =  \mdotst
,
\eeq
{\it i.e.} the mass flux through the inner radius is equal to the equilibrium 
flux. Therefore, this condition is equivalent to say that the inner disc 
instantaneously adapt itself to the conditions given by the outer disc. As 
discussed in PT99, this is consistent with the expression of the characteristic 
timescale as a function of the radius (Eq. \ref{eq_Tnu_disk}).

\subsection{Migration rate}

Dynamical tidal interactions of the growing protoplanet with the disc lead to 
two phenomena: inward migration and gap formation (Lin \& Papaloizou 1979, Ward 
1997, Tanaka {\it et al.} 2002). For low mass planets, the tidal interaction 
is linear, and migration is of type I (Ward 1997), whereas higher mass planets 
open a gap, leading to a reduction of the inward migration (referred to type 
II migration).

Analytical models of type I migration have been computed by Ward (1997). The 
resulting migration timescales are much shorter than both the disc lifetime 
and the planet growth timescale, making survival of forming planets unlikely: 
the planet is accreted onto the central star. Migration could be stopped if 
there is an inner cavity in the disc, but planets at larger distances remain
difficult to explain. Tanaka {\it et al.} (2002) have performed new analytical
calculations of type I migration, in two or three dimensional discs and found 
longer migration timescales but still too short to ensure survival. Their 
migration rate is nevertheless confirmed by recent three dimensional numerical 
calculations of disc structure and planet migration (Bate {\it et al.} 2003). 

On the other hand, suggestions of increased type I migration timescales can 
be found in Nelson \& Papaloizou (2004). As shown by these authors, the 
torques exerted on at least low mass planets ($\mplanet < 30 \mearth$) 
embedded in turbulent MHD discs are strongly fluctuating, resulting in a 
slowing down of the net inward motion. Contrary to laminar discs (as 
considered by Tanaka {\it et al.} 2002 and Bate {\it et al.} 2003) the 
migration proceeds as a random walk, and the mean value of the migration 
velocity seems to be highly reduced, compared to the laminar case. Moreover, 
as shown by Menou \& Goodman (2004), type I migration of low-mass planets 
can be slowed down by nearly one order of magnitude in regions of opacity 
transitions.

These considerations seem to indicate that the actual type I migration 
timescale may in fact be considerably longer than the one originally estimated 
by Ward (1997) or even by Tanaka {\it et al.}  (2002). For these reasons, 
and for lack of better knowledge, we actually use for type I migration the 
formula derived by Tanaka {\it et al.} (2002)  reduced by an arbitrary 
numerical factor $f_I$ chosen between $1/10$ and $1/100$. Tests have
shown that provided this factor is small enough to allow planet survival, 
its exact value \textit{does not} change the formation timescale but just 
the extent of the migration (see section \ref{sec_timescale}).

The migration velocity for low mass planets is taken to be: 
\beq
\dtotale{\aplanet}{t} = - 2 f_I \aplanet { \Gamma \over L_{\rm planet}}
\label{eq_tanaka}
,
\eeq
where $ L_{\rm planet} \equiv \Mplanet (G \Mstar \aplanet)^{1/2}$ is the 
angular momentum of the planet and the total torque $\Gamma$ is given by:
\beq
\Gamma = (1.364 + 0.541 \alpha_{\Sigma,P} )\( {\Mplanet \over \Mstar} { r_P \Omega_p \over C_{\rm s,P}} \)^2
\Sigma_{\rm P} r_P ^4 \Omega_p ^2
,
\eeq
where $C_{\rm s}$ is the sound velocity and $\alpha_\Sigma \equiv \dtotale{\log 
\Sigma}{\log r}$. In this expression, the subscript $P$ refers to quantities
at the location of the planet.

For type II migration, two cases have to be considered. For low mass planets 
(when their mass is negligible compared to the one of the disc) the inward 
velocity is given by the viscosity of the disc. As the mass of the planet 
grows and becomes comparable to the one of the disc, migration slows down and 
eventually stops. In this latter case, the variation of the planet orbital 
angular momentum is equal to the angular momentum transport rate in the 
gaseous disc (Lin {\it et al.} 1996, Ida \& Lin 2004):
\begin{equation}
{d \over d t} \left[ \mplanet \aplanet ^2 \Omega \right] = {3 \over 2} \Sigma \tilde \nu \Omega r^2
.
\end{equation}
In all cases of type II migration, the
migration rate is limited by the viscous transport in the disc:
\begin{equation}
{d a_{\rm planet} \over d t} = -{3 \nu \over 2 a_{\rm planet}} \times {\rm Min} (1,{2 \Sigma a_{\rm planet}^2 / M_{\rm planet}})
.
\end{equation}
Migration type switches from type I to type II when the planet becomes massive enough to
open a gap in the disc. It happens when the Hills radius of the planet becomes greater
than the density scale height $\tilde H$ of the disc.

\subsection{The planetesimals}

\subsubsection{Surface density and physical properties}
\label{surface_density_P}

The initial amount of heavy elements in the disc is a poorly constrained 
quantity.  For this reason, the dust-to-gas ratio is varied in our simulations, 
and takes two values depending on the mid-plane temperature of the disc: 
$f_{D/G}$ for temperatures below $150$ K and $1/4 f_{D/G}$ for higher 
temperatures. In principle, the position of the iceline should evolve due to 
the viscous evolution of the disc. However, since our treatment of the 
planetesimals disc is very simple, we do not take into account this evolution. 

We assume that due to the scattering effect of the planet, the surface density 
of planetesimals is constant within the current feeding zone but decreases 
with time proportionally to the mass accreted (and/or ejected from the disc) 
by the planet. The feeding zone is assumed to extend to a distance of 4 
$\rhill$ on each side of the planetary orbit, where $\rhill \equiv  \left({ 
\Mplanet  \over 3  \Mstar } \right)^{1 / 3} \aplanet$ is the Hills radius 
of the planet. For the inclinations and eccentricities of the planetesimals, 
we use the following prescription (P96):
\beq
i = {1 \over \aplanet} \sqrt{2 G M_{\rm planetesimal} \over r_{\rm planetesimal}} {1 \over \sqrt{3} \Omega}
,
\eeq
where $M_{\rm planetesimal}$ and $r_{\rm planetesimal}$ are the mass and radius 
of planetesimals, at the location of the planet, and 
\beq
e = {\rm max}(2 i, 2 {\rhill \over \aplanet})
.
\eeq

Finally, we also take into account the ejection of planetesimals due to 
the planet, using the  ejection rate  given by Ida \& Lin (2004):
\beq
{{\rm accretion \, rate } \over {\rm ejection \, rate }} = \( { V_{\rm esc,disk} \over V_{\rm surf,planet}} \)^4
\label{acc_to_ej_rate}
,
\eeq
where $V_{\rm esc,disk} = \sqrt{2 G \msun / \aplanet}$ is the escape velocity 
form the central star, at the location of the planet, $V_{\rm surf,planet} 
= \sqrt{G \mplanet / R_c}$ is the planet's characteristic surface speed, and 
$R_c$ is the planet's capture radius (see next section).

It is worth noting that our model for the evolution of the disc of
planetesimals remains a very simple one in which a number of effects are
neglected. For example, we omit the effect of gas drag on the planetesimals
which given their assumed size of $100$ km is reasonable. Using the
gas density obtained in our model, we can calculate the typical timescale
for radial drift. For regions above $\sim 4$ AU, we find values of order
of $\sim 10^6$ to $10^7$ the orbital time. In regions closer to the sun and
at the begining of our simulation, this timescale may be much lower owing
to the higher gas density. However, these inner regions evolve rapidly to
lower densities, and gas drag becomes quickly negligible.

We also neglect, apart for ejection and accretion onto the planet,
the perturbations created by the growing planet in the planetesimal disc such
as heating, gap formation and shepherding (Tanaka \& Ida 1997, Tanaka
\& Ida 1999, Thommes et al. 2003). For the latter effect, we have calculated
the critical migration rate, below which shepherding occurs (Tanaka \&
Ida 1999) and found that in all cases the migration rate exceeds this critical
value. The planet acts as a "predator" and not as a "shepherd" (see Tanaka \&
Ida 1999).

Apart from their radius, the planetesimals are characterized by a number of 
bulk properties like their density $\rho_{b}$, their tensile strength 
$\sigma_{T}$, heat of ablation (vaporization or melting) $Q_{abl}$ 
and some more material dependent parameters. This allows the simulation of 
the fate of different planetesimal types in the atmosphere of the growing 
protoplanet, as described in the next part.   

\subsubsection{Interaction with the growing atmosphere}

\label{infalling}

Given a core and the structure of the surrounding atmospheric envelope 
(pressure $P$, temperature $T$, density  $\rho$ {\it etc.} as a function of 
the distance from the planetary center), the second module computes the
trajectory and destruction of planetesimals inside this region by integrating 
a system of ordinary differential equations using a fifth-order Runge-Kutta 
method with automatic step size control (Press {\it et al.} 1992). This 
approach is similar to the one described by Podolak {\it et al.} (1988, 
hereafter referred to as PPR88). We will present this module and its results 
in more details in a oncoming separate paper (see Mordasini {\it et al.} 
2005), and restrict ourselves here to a description of the physical aspects 
we take into account. There are four main mechanisms controlling the fate 
of a planetesimal:

\textit{Gravity}: The gravitational attraction is calculated in the two-body 
approximation (planet - planetesimal) assuming a spherical mass distribution. 
This is justified because inside $\rhill$, where the calculation takes place,
the effect of the planet generally largely predominates other the third body 
effect. 

\textit{Aerodynamic drag force}: Apart from gravity, the aerodynamic drag 
force $F_D$ is the second force that defines the trajectory of the planetesimal. 
Using the standard formula for the drag force at high Reynolds numbers
(see e.g. Landau \& Lifshitz 1959) we have: 
\beq
F_D=\frac{1}{2}C_{D}\rho v^2 S
,
\eeq
 where $v$ is the velocity, $S$ is the instantaneous cross section of the 
planetesimal and $C_D$ is the drag coefficient (for a sphere), computed as a 
function of the local Mach and Reynolds numbers using the equations of Henderson 
(1976). These equations give smooth and continuous values for $C_D$ for widely 
varying aerodynamic environments (free molecular flow - continuum flow; 
hypersonic flow - incompressible flow).

\textit{Thermal mass loss}: The basic mechanism of thermal ablation is very 
simple: the drag force leads to a dissipation of kinetic energy, some fraction 
of which is used to heat the gas, while the remainder goes into heating up 
the body. When the heat flux is sufficiently high, the surface of the body 
becomes so hot that melting or vaporization starts. From these considerations, 
the simple classical ablation equation known from the study of terrestrial 
meteors is (\"{O}pik 1958, Bronsthen 1983):
\beq
\frac{dM}{dt}=-\frac{1}{2}C_{H}\rho v^3 S\frac{1}{Q_{abl}}
.
\eeq
 $Q_{abl}$ is the amount of energy per unit mass needed to bring body material 
from its initial temperature to the point where melting or vaporization occurs 
plus the specific heat needed for this phase change.  $C_H$, the heat transfer 
coefficient, is an unknown function which depends on the velocity of the 
particle, the flow regime, the shape of the body {\it etc.} and describes the 
fraction of the incoming kinetic energy flux of the gas that is available for 
ablation. 

One of the problem with the classical equation - apart from the neglect of 
re-radiation from the hot surface and heat conduction into the interior - 
is that the rate of thermal mass loss is heavily dependent upon the value of $C_H$
which can vary by several orders of magnitude depending upon the size of the infalling
body (from $10^{-5}$ to $\approx 0.5$, Svetsov {\it et al.} 1995).

For terrestrial, cm-sized meteorites a mean heat transfer coefficient of about 
0.1 can reasonably explain the observations (Bronsthen 1983). Whether this 
value can be extrapolated to impactors of tens or even hundreds of kilometers 
in size remained for a long time unclear (PPR88). Fortunately, the large number 
of studies of the impact of comet Shoemaker-Levy 9 (SL9) onto Jupiter have 
now shown that modelling such an impact with the above equation setting 
$C_H\approx 0.1$ greatly overestimates the mass loss and predicts too small 
a penetration depth in comparison to detailed hydrodynamic simulations 
(Sekanina 1993, Field \& Ferrara 1994, Ahrens {\it et al.} 1994). One can 
conclude from the SL9 event that for large, km-sized bodies, thermal ablation 
is of minor importance compared to mechanical destruction by aerodynamical 
forces (Svetsov 1995, see below) and that $C_H$ is small for such large objects, 
of the order of $10^{-4} - 10^{-3}$ (see e.g. Field and Ferrara 1995, Zahnle 
\& Mac Low 1994). 
Anyway, it is clear that thermal ablation must be considered in all cases 
even when fragmentation occurs, as it controls the final dissolution of the 
impactor (Korycansky {\it et al.} 2000), especially if one is interested 
not only in the energy deposition profiles (which was of major interest in 
the SL9 case) but also in the mass deposition (as it is the case here, since 
we aim to calculate the fraction of mass dissolved inside the atmosphere, and 
the actual mass of the solid core).

In order to obtain a realistic model of the energy input, we follow Zahnle 
(1992) by replacing the conductive heat influx term per surface unit $\propto 
\frac{1}{2}\rho v^3$ in the supersonic continuum flow by
$C_{H,rad}(T_{ps})\sigma T_{ps}^4$
whenever the latter is smaller, a way to take into account that in this regime, 
the energy input is proportional to the radiative flux from the shock front 
(\"{O}pik 1958). In this equation, $T_{ps}$ is the post-shock temperature
 which is directly computed as a function of the local atmospheric properties 
and the impactor velocity by solving numerically the normal shock wave jump 
conditions for a real gas (Landau \& Lifshitz 1959), using the equation of 
state of Chabrier {\it et al.} (1992). Our results for $T_{ps}$ are very 
similar to the ones obtained by Chevalier et Sarazin (1994). $C_{H,rad}(T_{ps})$ 
is a coefficient which denominates the fraction ($\leq1$) of radiation
reaching the planetesimal surface due the screening effect of ablated material. 
We use here the results of Bibermann {\it et al.} (1980).

We also include, where appropriate, radiation from the undisturbed atmosphere 
(PPR88) and a convective heat input mechanism which is proportional to the 
temperature difference between the surrounding gas and the impactor surface 
temperature. We use the expressions presented  by Sibulkin (1952), corrected 
-if necessary- for a turbulent boundary layer. A similar expression can also 
be found in Svetsov (1995).

The actual surface temperature of the planetesimal and the amount of ablated 
mass are obtained by equating the energy influx to the amount of re-radiation 
and energy used for ablation. For the reasons presented in PPR88 we neglect 
the heat conduction towards the interior of the planetesimal. For vaporization, 
$dM/dt$ is linked to the wall temperature via the Knudsen-Langmuir formula 
(Bronsthen 1983, PPR88).

The thermal mass loss rates found for large, non-fragmented bolides are in 
our simulation considerably lower than in a model using the simple ablation 
equation and $C_H=0.1$.

\textit{Mechanical destruction}: In deeper layers of the atmosphere, the 
aerodynamic pressure acting on the front side of the impactor can overcome 
the internal strength of the body (either tensile strength or self-gravity, 
see PPR88). The bolide will then approximately act as a fluid and undergo 
deformation, fragmentation and mechanical ablation (Svetsov {\it et al.} 
1995). Numerical simulations of the SL9 event have proven the prevalent 
importance of these effects to understand the atmospheric destruction of a 
large impactor (see e.g. Zahnle \& Mac Low 1994).  These effects occur in 
two different regimes. In the first one (static regime, when the aerodynamic 
pressure loading builds up on a time-scale larger than the time required for 
a sound wave to travel through the body, see Svetsov {\it et al.} 1995, 
Korykansky {\it et al.} 2000), we use a model which combines the effects 
of lateral spreading (using the so called ``pancake" model of Zahnle 1992) 
and the growth of Rayleigh-Taylor instabilities (Sharp 1984, Roulsten \& 
Ahrens 1997, Korycansky {\it et al.} 2000) at the front side of the fluidized 
impactor to obtain a multi-staged fragmentation model. In the second one 
(dynamical regime, when the pressure loading is dynamical), where no lateral 
spreading is observed (Svetsov {\it et al.} 1995,  Korycansky {\it et al.} 
2000) but where instead mass is removed at the front of the bolide, we use 
a simple model for mechanical ablation which is also based on the growth 
of RT instabilities.  For large impactors which are found in this regime, 
mass loss by mechanical ablation overcomes thermal ablation usually by a 
large factor.

After the calculation, the three dimensional energy and mass deposition 
profiles are converted back into one dimension to make them usable by our 
third module described in the next section. The planetesimal impact code is 
also used to calculate the capture radius, a calculation performed in a 
same way as described in P96.

\subsection{Protoplanet structure and evolution}

\subsubsection{Equations of the internal structure}

The third module calculates the planet internal structure including a growing 
core (including the energy deposited by accreted planetesimals) and the 
gas accretion due to both the contraction of the envelope and the increase of 
the outer radius of the planet. The standard equations of planet evolution 
are solved:
\begin{eqnarray}
{d r^3 \over d m} & = & {3 \over 4 \pi \rho} \label{eq_r} , \\
{d L \over d m}&  = & -{d U \over d t} + {P \over \rho^2} {d \rho \over d t} + \epsilon_{\rm acc} \label{eq_ener} , \\
{d P \over d m}&  = & {- G ( m + \mcore ) \over 4 \pi r^4} \label{eq_momentum}, \\
{d T \over d P}&  = & \nabla_{\rm ad} \,\,\, {\rm or} \,\,\, \nabla_{\rm rad} \label{eq_T} ,
\end{eqnarray}
using  opacities from Bell \& Lin (1994) and Alexander \& Fergusson (1994), 
and the equation of state (EOS) from Chabrier {\it et al.} (1992).  In these 
equations, $r,L,P,T$ are respectively the radius, the luminosity, the pressure 
and the temperature inside the atmosphere. These four quantities depend on the 
gas mass $m$ included in the sphere of radius $r$. $\rho$ is the mass density, 
$U$ the specific internal energy, and $\mcore$ the mass of the solid core.
$\epsilon_{\rm acc}$ is the amount of energy released by the accretion of 
planetesimals (see section \ref{infalling}); this term  dominates largely the 
energy budget in Eq.  \ref{eq_ener}, until the maximum accretion rate has been 
attained (see Section \ref{mdotgas}).  The temperature gradient is given by the 
adiabatic ($\nabla_{\rm ad}$) or by the radiative gradient ($\nabla_{\rm rad}$),
depending on the stability of the zone against convection which we check using 
the Schwarzschild criterion. For convective zones, we assume that the temperature 
gradient is given by the adiabatic one; in other words, we do not use the mixing 
length theory (MLT) in these models. Test have shown that including MLT does 
not change our models.

Two approximations can be used regarding the fate of planetesimals' matter,  
after their destruction inside the envelope. In the "sinking" approximation 
(see P96), this mass is assumed to slowly sink toward the core, leading to an 
extra term in the core luminosity. In the "no sinking" approximation, the matter 
is assumed to remain inside the envelope. Note that, as in P96, the matter 
is added to the core mass in Eq. \ref{eq_momentum} in both cases.  Under this 
aspect, the sinking approximation appears to be more self-consistent.

Two inner boundary conditions are necessary to solve Eq. \ref{eq_r} to \ref{eq_T}:
the core radius is set to $R_{\rm core} =\left( { M_{\rm core} \over 4/3 \pi 
\rho_{\rm core} } \right)^{1/3}$ and the core luminosity  $L_{\rm core}$ is 
given by the sum of the remaining kinetic plus the corresponding potential
energy of planetesimals, after having passed through the atmosphere. The 
density of the core is fixed to $3.2$ g/cm$^3$ as in P96.

The  mass of the envelope is given by requiring that the outer radius of the planet
is equal to  the minimum of the Hill radius and the accretion radius $R_{\rm 
accr} \equiv { G \Mplanet \over C_s^2 }$ where $C_s$ is the sound velocity 
inside the disc at the location of the planet (see P96).

Finally, two outer boundary conditions are necessary. They are given by 
requiring that the disc and the planet join smoothly at the outer radius:
\beq
P_{\rm surf} = P_{\rm mid}(\Sigma(\aplanet,t))
\eeq
and
\beq
T_{\rm surf} = T_{\rm mid}(\Sigma(\aplanet,t))
,
\eeq
where $P_{\rm mid}(\Sigma(\aplanet,t))$ and $T_{\rm mid}(\Sigma(\aplanet,t))$ 
are calculated using the structure of the disc (see section \ref{disk_vert}).

This condition is valid as far as the disc can supply enough mass to keep the 
outer radius equal to the Hill (or the accretion) radius, {\it i.e.} if the 
gas accretion rate is below the maximum accretion rate (see section \ref{mdotcore}).

\subsubsection{Solid accretion rate}
\label{mdotcore}

We  use the expression of Greenzweig \& Lissauer (1992) to calculate the solid 
accretion rate:
\beq
\dtotale{M_{\rm solids}}{t} = \pi R_c^2 \sigmaP \Omega F_g
,
\eeq
where $\sigmaP$ is the surface density of solids at $r=\aplanet$ and $R_c$ is 
the capture radius of the planet. The capture radius is calculated using our 
second module (see section \ref{infalling}).

\subsubsection{Gas accretion rate}
\label{mdotgas}

Normally, the gas accretion rate is determined from the condition:
\beq
R(m=\Mplanet) = {\rm min}(\rhill,\racc)
,
\eeq
where $\racc$ is the accretion radius of the planet (see P96).
At each timestep, the mass of the envelope (and then the total mass) is 
determined by iterations to satisfy this condition to a given accuracy. 
From comparing models at $t$ and $t+dt$, we obtain the gas accretion rate.

The condition $R(m=\Mplanet) = {\rm min}(\rhill,\racc)$ can be fulfilled
only if the disc can supply enough gas to the planet.  Once a gap opens in
the disc, the maximum gas accretion rate is set to the rate given by (Veras \&
Armitage 2004):
\beq
{\dot{M}_{\rm gas,max} \over \dot{M}_{\rm disc} }= 1.668 \left( \Mplanet / M_J \right)^{1/3}
e^{- {\Mplanet \over 1.5 M_J}} + 0.04
\label{eq_max}
,
\eeq
where $M_J$ is the Jupiter mass and 
$\dot{M}_{\rm disc}$
is the disc accretion rate away from the planet.  When the maximum accretion rate 
is reached, the growth of the planet  mass is set by the disc and no longer 
by the internal structure of the planet, which is no longer computed.

\subsection{Initial conditions and physical assumptions}

We start our calculations with a core of $0.6 \mearth$, at a distance $\aini$ 
from the star. The initial surface density of the disc is usually taken as a 
power law, the total mass being given (for fixed boundaries) by the normalisation 
at $5.2$ AU. Its life time is given by the value of $\alpha$ and the rate of 
photo-evaporation.  At last, the solid surface density is fully characterized 
by the dust to gas ratio $f_{D/G}$.

In addition, two numerical parameters have to be specified, $f_I$ (reduction 
of type I migration) and $f_\Lambda$, the numerical factor in the expression 
of the momentum transfer between the planet and the disc.  As  shown in  
Alibert {\it et al.} (2004), modifications in the outer planet boundary 
conditions due to the gap formation have a small effect.  Therefore, the 
gap essentially limits the gas accretion onto the planet, and we will 
restrict ourselves to the case $f_\Lambda = 0$.

Some other parameters of the model, which are not changed in the calculations 
presented here are summarized in Table \ref{table_para}.

\begin{table}
\caption{Physical parameters used in our models}
\begin{tabular}{lccc}
\hline\noalign{\smallskip}
Parameter & Value \\
\noalign{\smallskip}
\hline\noalign{\smallskip}
Core density   & $3.2$ g/cm$^3$ \\
Initial core mass & $0.6 \mearth$ \\
Helium fraction & $0.24$ \\
Internal disc radius & $0.25$ AU \\
External disc radius & $50$ AU \\
Planetesimal radius & $100$ km \\
\hline
\end{tabular}
\label{table_para}
\end{table}

\subsection{Numerical tests}

\begin{figure}
\begin{center}
\epsfig{file=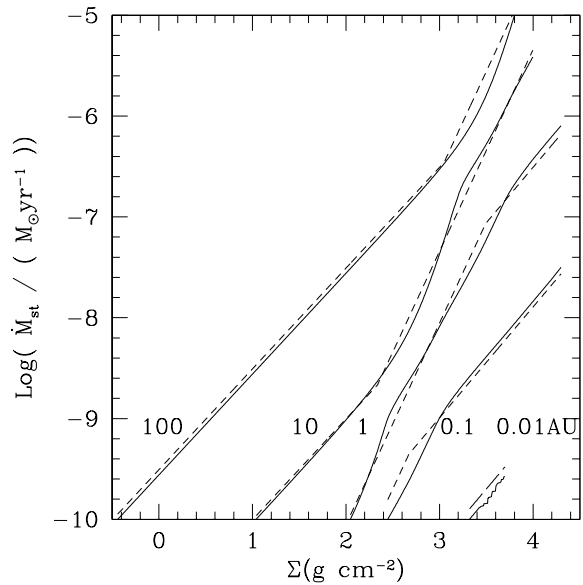,height=80mm,width=80mm}
\end{center}
\caption{$\mdotst(r,\Sigma)$ given by our calculations (solid lines), and the 
analytical fits by PT99 (dotted lines), for $\alpha = 10^{-3}$. The values 
of the radii (in AU) at which these rates are computed are indicated on the plot.
}
\label{comp_PT99_disk}
\end{figure}

Various tests have been performed to validate our entire model.
Concerning the disc part of our code, we have compared the resulting 
functions $\tilde{\nu}(r,\Sigma)$ for different values of $\alpha$ to the 
analytical fits given in PT99. Figure \ref{comp_PT99_disk} shows a comparison 
for $\alpha=10^{-3}$. We recall that PT99 mention a difference between the 
numerical results and their fits of the order of $50$\% at most. This is 
also comparable to the maximum difference between our results and the fits.

\begin{figure}
\begin{center}
\epsfig{file=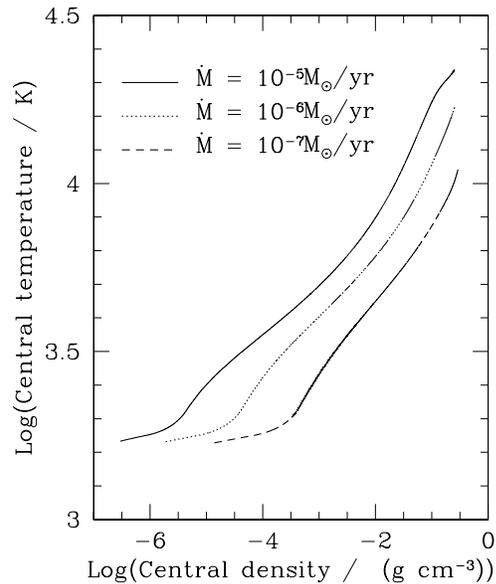,height=80mm,width=80mm}
\end{center}
\caption{Central density and temperature for models analog to case 1, 6 and 7 of BP86.
}
\label{compBP86}
\end{figure}

The interior structure module has been used to reproduce the results
of BP86. Figure \ref{compBP86} shows the evolution of a forming Jupiter, for a 
planet below the cross-over mass (until the mass of accreted planetesimals and
the mass of accreted gas are equal).  
In this calculation, the accretion rate of planetesimals is 
constant, and there is no interaction between the envelope and the planetesimals. 
The boundary conditions are those of case 1, 6 and 7 of BP86. Figure \ref{compBP86} 
shows the central temperature and density for these models, which is in good 
agreement with their results (see Figure 1 in BP86). We conclude that our 
internal structure module works properly.

\begin{figure}
\begin{center}
\epsfig{file=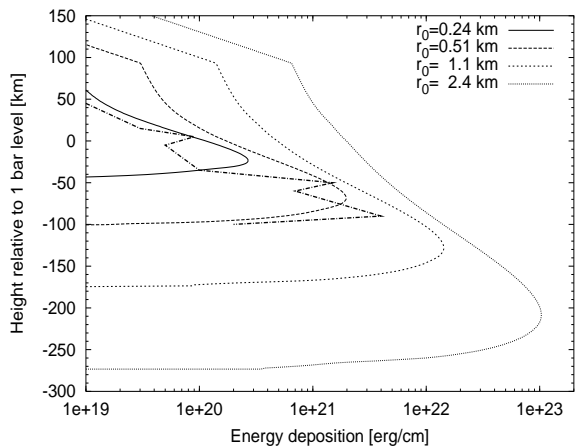,height=60mm,width=80mm}
\end{center}
\caption{Energy deposition profiles of SL9-type impactors with different initial 
radii in the Jovian atmosphere. For comparison, a schematic representation of the 
energy deposition profile of a 0.5 km impactor in the high-resolution hydrodynamic 
simulations of Mac Low \& Zahnle (1994) is given (thick line).
}
\label{compSL9}
\end{figure}

Our planetesimal accretion module has been tested extensively by comparing results 
of simulations of impacts into terrestrial and Venusian atmosphere (ReVelle 
1978; Hills \& Goda \cite{hillsgoda93}, Zahnle 1992), as well as into Jupiter's 
atmosphere (Zahnle \& Mac Low 1994). Figure \ref{compSL9} shows one of these 
tests: energy deposition profiles are plotted for compact ice comets of 
different initial radii hitting the Jovian atmosphere. The initial conditions 
are chosen to match the SL9 event. As expected, larger bolides penetrate 
deeper and have a larger peak energy deposition. Our results are virtually 
identical to those of the analytical model of  Zahnle \& Mac Low (1994), which 
is not surprising as the pancake model is used in both cases. The edge at an 
altitude of about 90 km comes from the radiative limit for thermal ablation and 
appears at the same point as in Zahnle \& Mac Low (1994) and Crawford (1996). 
The thick line schematically represents an energy release profile obtained from  
a high-resolution hydrodynamic simulation by Mac Low \& Zahnle (1994), 
predicting a very similar peak energy deposition altitude and  value. A model 
using the classical ablation equation and a high heat transfer coefficient 
($C_H \sim 0.1$), but no mechanical effects can not reproduce this profile.

\begin{figure}
\begin{center}
\epsfig{file=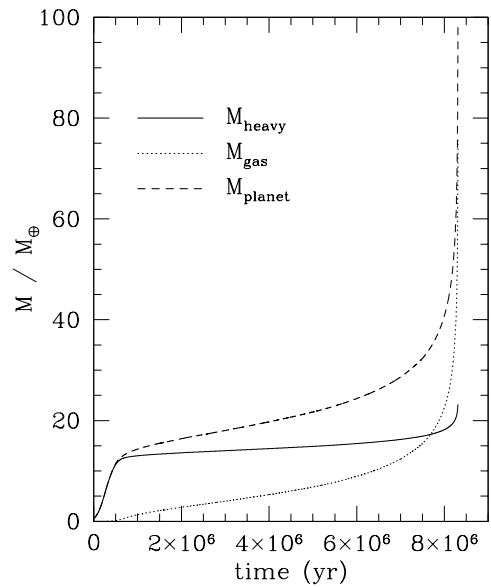,height=80mm,width=80mm}
\end{center}
\caption{Mass of accreted planetesimals ($M_{\rm heavy}$, solid line), mass of 
accreted gas (dotted line), and total mass of the planet (dashed line) as a 
function of time, for conditions similar to case J1 of P96.
}
\label{J1}
\end{figure}

Finally, the entire code has been tested using the same initial conditions as 
P96 (case J1) turning disc evolution and migration off. In this case, we obtain 
a cross-over time of $\sim 8$ Myr, in close agreement with their result 
(compare Fig. \ref{J1} with Fig. 1 of P96). We conclude that our code properly 
reproduces giant planet formation in absence of migration and disc evolution. 

\section{Results}
\subsection{Formation timescales}
\label{sec_timescale}

As shown in BP86 and P96, the formation timescale is essentially given by
the time needed to reach the cross-over mass. 
To quantify the effects of migration and disc evolution 
on planet formation, we compare different models, all reaching the cross-over 
phase at the same location in the disc, namely $5.5$ AU. 

For comparison purpose with P96, we consider an initial disc density profile 
given by $\Sigma \propto r^{-2}$, like in  P96, again for comparison purpose. 
The constant is chosen to yield $\Sigma = 525$ g/cm$^2$ at 5.2 AU; this 
corresponds to $\sim 3$ times the surface density in the minimum mass solar 
nebula at the position of Jupiter. This surface density profile yields 
isolation masses that do not depend on the distance to the sun. We do not 
take into account photo-evaporation, and we start with an embryo of 0.6 
$\mearth$ initially at 7, 8 or 15 AU, depending upon the choice of the parameter 
$f_I$ (0.01, 0.03 and 0.1 respectively).  The viscosity parameter $\alpha$ 
is set to $2.10^{-3}$ and the dust-to-gas ratio is equal to $1/70$ for disc 
mid-plane temperature below 150 K, and $1/280$ in the opposite case.  Finally, 
we use for these simulations the no-sinking approximation (see P96).

\begin{figure}
\begin{center}
\epsfig{file=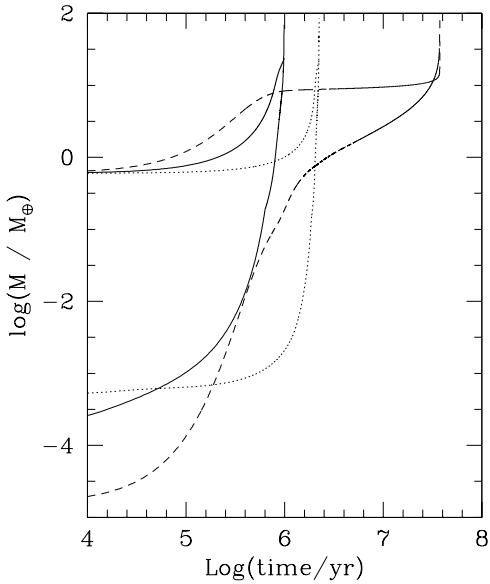,height=80mm,width=80mm}
\end{center}
\caption{Mass of accreted planetesimals (lines starting at $0.6 \mearth$), 
and mass of accreted gas as a function of time, until the cross-over mass
is reached. Solid lines: starting at 8 AU (with $f_I = 0.03$), dotted lines: 
starting at 15 AU (with $f_I = 0.1$) and dashed lines: {\it in situ} model. 
The model starting at 7 AU is very close to the one starting at 8 AU and is 
not shown.}
\label{timescale}
\end{figure}

Figure \ref{timescale} shows the mass of accreted planetesimals and gas as a 
function of time for the models considered here.  Note that the mass of accreted 
planetesimals does not correspond to the core mass since some fraction of 
them are being destroyed while traversing the envelope and never reach the 
core. For the {\it in situ} case (without disc evolution), the time to reach 
the cross-over mass is around 30 Myr, much longer than the typical discs 
lifetimes (Haisch et al. 2001). For the simulations with migration and disc 
evolution, the corresponding timescales are respectively $\sim 1$, $\sim 1$ 
and $\sim 2.2$ Myr for an embryo starting at 7, 8 or 15 AU respectively.

Allowing for migration and disc evolution, we obtain a time to reach the 
cross-over mass of about ~1-3 Myr, {\it i.e.} more than ten times faster 
than in our identical model in which migration and disc evolution have been 
switched off. As explained in Alibert {\it et al.} (2004), the main reason for 
this speed-up is found to be due to migration which prevents the severe 
depletion of the feeding zone as observed in the {\it in situ} formation 
model. Migration therefore suppresses the need to accrete a gas envelope 
in order to extend the feeding zone (phase 2 of P96) and the planet reaches 
cross-over mass much faster.

\begin{figure}
\begin{center}
\epsfig{file=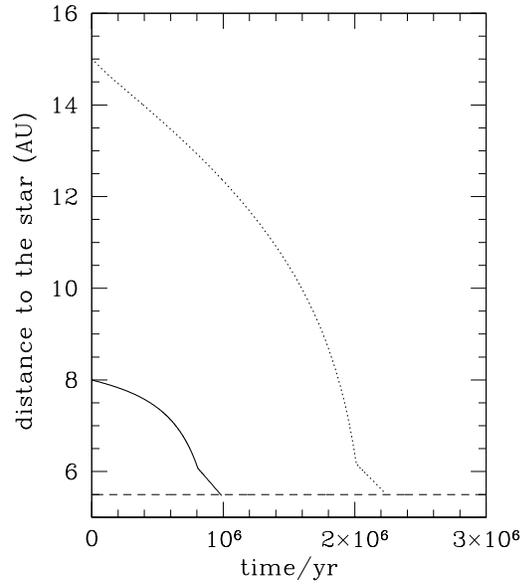,height=80mm,width=80mm}
\end{center}
\caption{Distance to the central star for the same models as in figure 
\ref{timescale}. The kinks around 0.8 and 2 Myr signal the change from type I 
to type II migrations. Solid line: starting at 8 AU (with $f_I = 0.03$), dotted 
line: starting at 15 AU (with $f_I = 0.1$), and dashed line: {\it in situ} model.
}
\label{distance}
\end{figure}

As stated in section \ref{infalling}, the effect of ablation is found to be 
negligible, disruption of planetesimals occurring very deep in the envelope.  The 
two cases, sinking and no-sinking, considered by P96 differ therefore much less 
in our calculations than in P96. For the simulation started at 8 AU, the time 
to reach the cross over mass is found to be of order $\sim 1.2$ Myr using the 
sinking approximation, compared to $\sim 1$ Myr in the no-sinking case.  
For that reason, and since it is more self-consistent, we will use in the 
following the sinking approximation, as in P96's standard calculations.

\subsection{Formation of extrasolar planets}
\label{warm_J}

\begin{figure}
\begin{center}
\epsfig{file=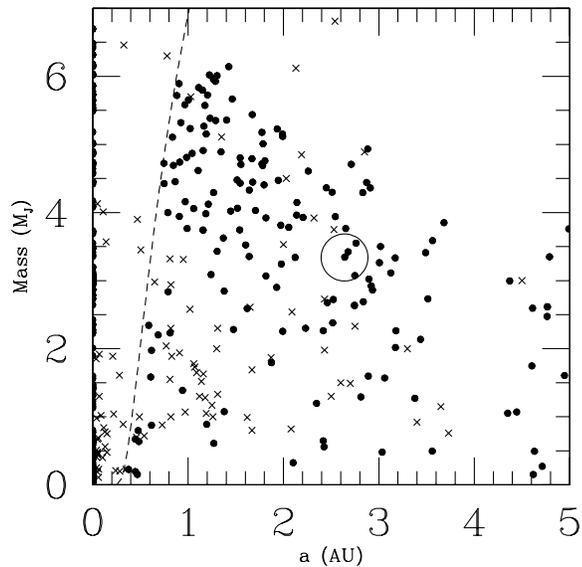,height=80mm,width=80mm}
\end{center}
\caption{Final mass and semi-major axis (dots), for a set of simulations,
using different initial parameters (see text for details).  The crosses are 
the parameters of the observed exoplanets as in November 2003.  The big 
circle represents the final parameters of the model detailed in section 
\ref{warm_J}. The dashed line gives the inner limit of our simulations 
(when the inner radius of the planet feeding zone is below 0.25 AU).}
\label{m_a}
\end{figure}

Depending on the initial parameters used, very different planets can form.
Figure \ref{m_a} shows the results of a set of 1000 simulations\footnote{Each simulation takes $\sim 3$ to 4
hours on a modern PC.}, 
in a mass {\it vs} semi-major axis diagram. The initial parameters for these simulations
are $\aini = 3$, 4, 5, 6, 7.5 10, 15 and 20 AU, $f_{D/G} = 1/25$, $1/50$, $1/70$, $1/100$ and $1/200$,
and a photoevaporation
rate of $10^{-9} \msun/$yr, $2 \times 10^{-9} \msun/$yr, $4 \times 10^{-9} \msun/$yr, $10^{-8} \msun/$yr
and $2 \times 10^{-8} \msun/$yr.
The initial disc is given by $\Sigma \propto r^{-3/2}$, the normalization at
$5.2$ AU being between 100, 300, 500, 700 and 900 ${\rm g/cm^2}$. This 
corresponds to disc masses of $\sim 0.01$ to $\sim 0.1 \msol$ between 0.25 
and 50 AU.  The viscosity parameter is fixed to $2 \times 10^{-3}$, and we 
use $f_I = 0.03$ for the whole set of simulations.  The edges of the disc 
are at 0.25 and 50 AU, so planets whose inner radius of the feeding zone 
is below 0.25 AU are considered to have migrated inside the star, and are 
represented by dots on the vertical axis.

One must keep in mind that the diagram shown here {\it does not} take into 
account any statistical weighting of the different initial parameters, so 
one should be careful when comparing the obtained distribution with the 
observed one. We present it simply to illustrate the possible diversity of 
planets that can form in our model. In a forthcoming paper we will discuss 
the probability of occurrence of these planets.

\begin{figure}
\begin{center}
\epsfig{file=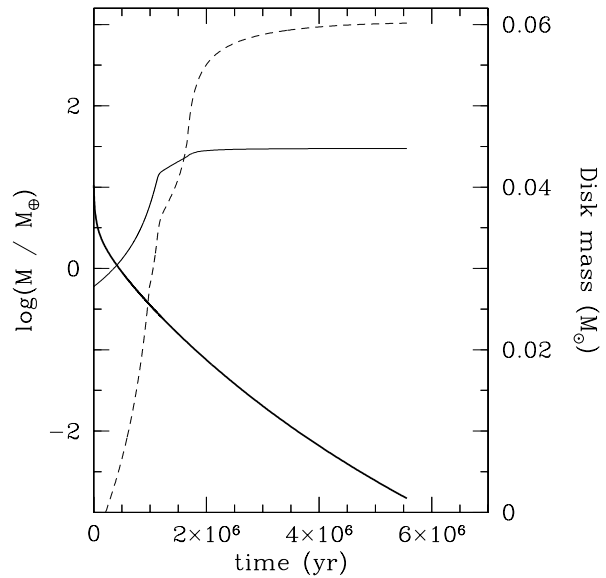,height=80mm,width=80mm}
\end{center}
\caption{Mass of the different components for model of section \ref{warm_J}. 
Solid line: mass of accreted planetesimals, dashed line: mass of H/He, and 
heavy solid line: mass of the disc.}
\label{Mass_warm_J}
\end{figure}

\begin{figure}
\begin{center}
\epsfig{file=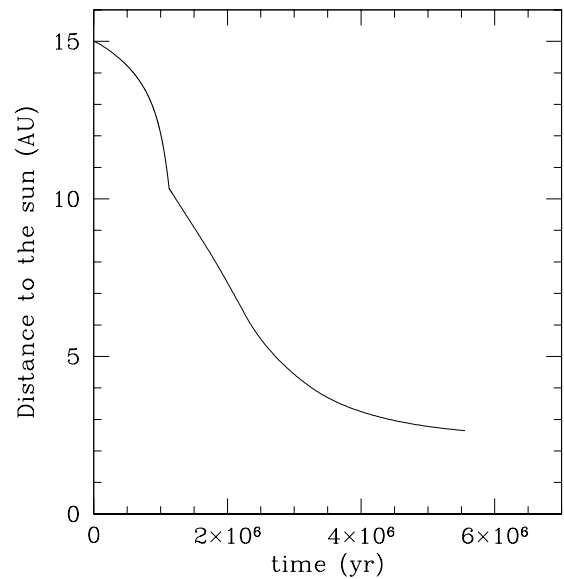,height=80mm,width=80mm}
\end{center}
\caption{Distance to the sun for model of section \ref{warm_J}. The kink around 1 Myr 
corresponds to the change from type I to type II migration.}
\label{distance_warm_J}
\end{figure}

As an illustration taken from this set of simulations, we provide some 
details on the formation process of one particular planet. This object 
evolved from the following: $\aini = 15$ AU, a disc photo-evaporation rate 
$4 \times 10^{-9} \msun/$yr, an initial normalisation of the disc at 5.2 
AU equal to 500 g/cm$^2$, and $f_{D/G} = 1/70$.  With a final mass of 
$\sim 3.5 M_J$, and a final distance $\sim 2.5$ AU from its star, this 
planet is indicated by the circle in Figure \ref{m_a}.  Figure \ref{Mass_warm_J} 
shows the evolution of the mass of accreted gas, the mass of 
accreted planetesimals, as well as the mass of the disc, and Figure 
\ref{distance_warm_J} gives the distance to the central star as a function 
of time. The cross-over mass is reached after $\sim 1.6$ Myr, and shortly 
after, due to gap formation, the accretion rate of gas is limited to its 
maximum value, which decreases with decreasing disc mass.  The formation 
process ends after 5.5 Myr when the disc has essentially disappeared. 
Note however, that $90 \%$ of final mass of the planet is reached already 
after $\sim 4$ Myr. At the end of the simulation, the planet as accreted 
$\sim 30 \mearth$ of planetesimals.  As stated in Section
\ref{sec_timescale}, this mass does not correspond to the mass of the core,
since some of the accreted planetesimals may have been destroyed during
the crossing of the envelope. Our model provides an estimate of the core
mass by tracking the location where the planetesimals are destroyed (second
module). For this simulation, $\sim 6.4 \mearth$ of accreted planetesimals 
reach the core without being destroyed in the envelope. Ignoring further
processes, this provides the mass of the core which is similar to the one 
inferred from Jupiter's internal structure models (Guillot et al. 2004).
The total content in heavy elements will finally depend on the metallicity
of the accreted gas. For solar composition (prior to the formation of 
planetesimals), the accreted gas add an additional $\sim 7 \mearth$ of 
heavy elements.

The migration of the planet can be divided into three phases. Before 
$\sim 1$ Myr the planet undergoes type I migration at which time a gap 
opens and migration switches to type II. Shortly after $\sim 2$ Myr, the 
mass of the planet becomes non negligible compared with the disc mass and 
migration slows down and eventually stops at the time the disc disappears.
 
\section{Summary and discussion}

We have presented a new code devoted to the calculation of giant planet formation
models, taking into account the protoplanetary disc structure, as well as the 
migration of the forming planet. These calculations show that the formation of 
giant planets, at least in the first phase until runaway gas accretion, can be 
heavily sped-up if one takes into account the effect of migration. This is 
mainly due to the suppression of phase 2 as described in P96.  Using an initial
disc model similar to the one of P96, we obtain a time to reach the cross-over 
mass of the order of 1 Myr, significantly shorter than the typical disc lifetimes.
Moreover, this speed-up due to migration has been found to be robust against 
reasonable changes in various parameters (Alibert {\it et al.} 2004). Therefore,
migration not only accounts for the orbital distribution of extra-solar planets,  
but also considerably shortens the formation timescale in the core-accretion 
scenario. The formation of giant planets in this scenario is therefore in 
excellent agreement with inferred disc lifetimes without having to consider 
discs significantly more massive than the minimum mass solar nebula.

We note that the speed-up due to migration does not preclude other 
effects to further reduce this timescale. For example, reducing the dust 
opacity would decrease the critical mass (Ikoma et al. 2001), thus leading 
to another reduction in the formation timescale. Furthermore, such a reduced 
opacity could account for the existence of giant planets with small central 
core.

Using different initial conditions can lead to the formation of a wide variety 
of planets.  However, the comparison of our results with observations of 
extrasolar planets need to take into account the probability distribution of 
various initial conditions.  Work is under progress to obtain, using a 
Monte-Carlo approach, synthetic distributions which can then be compared in a 
statistical way with the observed distributions.

These calculations are however subject to some uncertainties, among them
the simplified treatment of the planetesimals disc, or the calculation of
the ejection rate, that directly determines  the final heavy elements content.
Work is under progress to improve these aspects.

Finally, it seems to be very difficult to form a planet, and to prevent it 
from spiraling into the sun if the amount of type I migration as computed 
today is not reduced by a factor of at least 10. 

\acknowledgements

This work was supported in part by the Swiss National Science Foundation.

\end{document}